\documentstyle[aas2pp4]{article}

\def\bref{\par\noindent\hangindent 20pt}
\def\J16{{GRO J1655--40}}
\def\Msun{\rm M_{\odot}}

\begin{document}

\title{Advection-Dominated Flows and the X-ray Delay in the Outburst of GRO
J1655--40}

\author{J.-M. Hameury}
\affil{URA 1280 du CNRS, Observatoire de Strasbourg,11 rue de l'Universit\'e, 
F-67000 Strasbourg, France}
\author{J.-P. Lasota}
\affil{UPR 176 du CNRS; DARC, Observatoire de Paris, Section de Meudon, 
92195 Meudon, France}
\and
\author{J.E. McClintock and R. Narayan}
\affil{Harvard-Smithsonian Center for Astrophysics, 60 Garden Street, 
Cambridge, MA 02138, USA}

\begin{abstract}

We show that the time delay between the optical and X-ray outbursts of the
black-hole soft X-ray transient source GRO J1655--40, observed in April 1996,
requires that the accretion flow in this object must consist of two
components: a cold outer accretion disk and an extremely hot inner
advection-dominated accretion flow (ADAF). In quiescence, the model predicts
a spectrum which is in good agreement with observations, with most of the
observed radiation coming from the ADAF. By fitting the observed spectrum,
we estimate the mass accretion rate of the quiescent system and the
transition radius between the disk and the ADAF. We present a detailed
numerical simulation of a dwarf-nova type instability in the outer disk. The
resulting heat front reaches the ADAF cavity promptly; however, it must then
propagate inward slowly on a viscous time scale, thereby delaying the onset
of the X-ray flux. The model reproduces the observed optical and X-ray light
curves of the April 1996 outburst, as well as the 6--day X-ray delay.
Further, the model gives an independent estimate of the quiescent mass
accretion rate which is in very good agreement with the rate estimated from
fitting the quiescent spectrum. We show that a pure thin disk model without
an ADAF zone requires significant tuning to explain the X-ray delay;
moreover, such a model does not explain the quiescent X-ray emission of GRO
J1655--40.

\end{abstract}


\section{Introduction}

The binary X-ray source GRO J1655--40 (also called X-ray Nova Scorpii 1994)
is a member of the class of so-called ``Soft X-ray Transients'' (SXTs) or
``X-ray Novae''. In these systems, a low mass Roche-lobe-filling secondary
star transfers mass through an accretion disk onto a compact object: a
neutron star or a black hole. Compared to a neutron-star transient, a
black-hole transient (BHT) generally has a larger X-ray outburst amplitude
and a lower quiescent luminosity, which is a signature of a black hole's
event horizon (Narayan, Garcia \& McClintock 1997). The mass of the black
hole primary in GRO J1655--40 is $\approx 7.0 \Msun$ (Orosz \& Bailyn 1997;
hereafter OB).

GRO J1655--40 is an exceptional BHT because of its frequent outbursts in
recent years. Most BHTs have recurrence times of decades or longer, whereas
GRO~J1655--40 has gone into outburst several times since its discovery on
1994 July 27 by BATSE on the {\sl Compton Gamma Ray Observatory} (CGRO)
(Zhang et al. 1994). Two subsequent outbursts occurred in late March 1995
(Wilson et al. 1995) and in July 1995 (Harmon et al. 1995). Following an
extended period of X-ray quiescence, the source again went into outburst in
April 1996, as discovered by the All-Sky Monitor (ASM) on the {\sl Rossi
X-ray Timing Explorer} (RXTE) (Remillard et al. 1996; Levine et al. 1996).

Thus GRO J1655--40 has remained active off and on for nearly three years.
Other BHTs have shown X-ray and optical activity several months after an
outburst; however, none of them have sustained their activity for more than
about a year (e.g. Tanaka \& Shibazaki 1996). The frequent outbursts of GRO
J1655--40 in recent years may be due to an enhancement of the mass transfer
rate, which is estimated to be relatively high, $2.2 \times 10^{17}$ g
s$^{-1}$ (OB). One should however note (Ritter 1997) that this estimate is
based on a formula of King, Kolb \& Burderi (1996) which is valid for giants,
but not for systems such as \J16 which have not yet reached the giant branch.
Systems with companions in the Hertzsprung gap should transfer mass at a very
high rate ($\gtrsim 10^{19}$ g s$^{-1}$, Ritter 1997), since the secondary
expands on a thermal time; this is obviously not the case now in GRO
J1655--40, and no observational determination of the mass transfer rate is
available at present. As it happens, the value given by OB is plausible, and
we shall use it in the following. In any case, on longer timescales GRO
J1655--40 behaves more like other BHTs, since no outburst of the source has
been reported in the previous 25 years. GRO J1655--40 is also distinguished
by its radio outbursts, which are associated with superluminal expansion
events (Tingay et al. 1995; Hjellming \& Rupen 1995).

\begin{figure}
\plotone{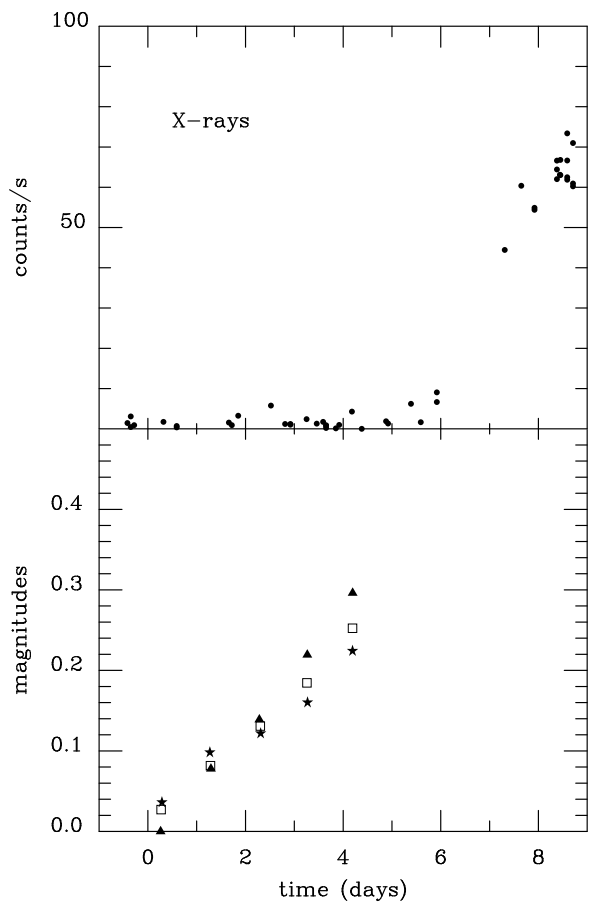}
\caption{Observed optical and 2--12 keV X-ray light curves during
the initial phase of the April outburst of \J16 (From ORBM). For the sake of
clarity, only one average data point per night is plotted. The B, V and I
magnitudes are represented by triangles, squares and stars respectively. Time
has been set to zero at an arbitrary point close to the onset of the
outburst.}
\end{figure}

About 6 days prior to the most recent X-ray outburst of GRO J1655--40 (in
April 1996), a remarkable optical precursor was observed (Orosz et al. 1997;
hereafter ORBM). As shown in Figure 1, starting from an initially quiescent
state, the optical intensities (BVRI) were observed to rise gradually for
several days and brighten by about 30\% before the onset of the X-ray
outburst. In this article, we examine only one aspect of the complex
behavior of GRO J1655--40 in outburst, namely, the properties of the optical
precursor and the X-ray outburst, and what they imply for models of quiescent
BHTs and for the outburst mechanism. As argued by ORBM, the substantial
delay between the optical eruption and the X-ray outburst (which we refer to
hereafter as the ``X-ray delay'') may provide support for the
advection-dominated accretion flow (ADAF) model of the inner regions of the
quiescent accretion disk.

One outburst mechanism that has been developed for SXTs is the mass transfer
instability (Hameury, King \& Lasota 1986); however, in general it cannot
reproduce the characteristic timescales of SXTs and it has therefore been
rejected (Gontikakis \& Hameury 1993). It is now clear that the outburst
mechanism must be due to a disk instability. A natural candidate for such an
instability is the thermal (and viscous) instability resulting from abrupt
changes in opacities when hydrogen becomes partially ionized. Such a
mechanism explains successfully dwarf nova outbursts in the framework of the
disk instability model (DIM) (see Cannizzo 1993 and references therein). The
dwarf nova DIM has been extended to SXTs by Mineshige \& Wheeler (1989) (see
also Cannizzo, Chen \& Livio 1995).

The DIM requires the quiescent accretion disk to be in a cold state; the
accretion rate therefore has to be everywhere lower than a critical accretion
rate $\dot M_{\rm crit}(R) \propto R^{\rm a}$ where $a \approx 2.6$ (see e.g.
Ludwig, Meyer-Hofmeister \& Ritter 1994). This in turn implies that the
accretion rate onto the compact object required by the DIM is extremely low
$\sim 10^6$ g s$^{-1}$ (Mineshige \& Wheeler 1989; Lasota 1996a). However,
observations of X-ray emission from quiescent BHTs imply accretion rates
which are several orders of magnitude higher (McClintock, Horne \& Remillard
1995; Verbunt 1996; Narayan, McClintock \& Yi 1996 (NMY); Narayan, Barret \&
McClintock 1997 (NBM); Robinson et al. 1997). A similar problem is
encountered in some quiescent dwarf novae (see e.g. Lasota 1997). It is clear
therefore that the standard DIM cannot apply to BHTs (Lasota 1996a,b);
furthermore, in some cases it must be modified even to describe dwarf-nova
outbursts (Meyer \& Meyer-Hofmeister 1994; Livio \& Pringle 1992).

A model for SXTs in quiescence was proposed by NMY in which the accretion
flow occurs as a thin disk only outside a transition radius $R_{\rm tr} \sim
10^{3}$ Schwarzschild radii, while for $R<R_{\rm tr}$ the flow forms an
advection-dominated accretion flow (ADAF) (Abramowicz et al. 1995; Narayan
\& Yi 1994, 1995; for a recent review see Narayan 1997). In the NMY model,
the observed X-rays are emitted with a very low efficiency by the ADAF while
the UV and optical luminosity is produced by the outer disk. However,
Lasota, Narayan \& Yi (1996) pointed out that the NMY model is not self
consistent because the outer disk is relatively hot and therefore subject to
a thermal instability, contrary to the assumed stationarity (see also Wheeler
1996). More recently, NBM have shown that self-consistent models for the
spectra of V404 Cyg and A0620--00 can be obtained by an ADAF that extends
outward to $R_{\rm tr} \sim 10^{4}$ Schwarzschild radii. For these models,
the accretion disk is cool and the optical/UV flux is mostly supplied by
synchrotron emission from the ADAF. In such a model, the transient outburst
originates in the outer cold disk and is due to a dwarf-nova type instability.

In section 2, we discuss the consequences of the delay between optical and
X-ray, and in section 3 we develop an ADAF model for GRO J1655--40 in
quiescence. In section~4 we demonstrate that a time-dependent model of the
outburst of GRO~J1655--40 implies the existence of a two-component accretion
flow. The parameters determined for this flow agree with the parameters
independently determined in section~3.

\section{Interpretation of the X-ray Delay: Evidence for a Two-Component
Disk}

\subsection{The UV-Delay in Dwarf-Nova Outbursts}

The X-ray delay observed in the outburst of GRO J1655--40 is analogous to the
well known UV delay observed for dwarf novae (e.g. Warner 1995, and
references therein). For dwarf novae the rise in the UV flux starts about 5
to 15 hours after the beginning of the optical outburst. In the framework of
the standard disk instability model (DIM) one can interpret the UV-delay as
due to an ``outside-in'' (or Type A) outburst. According to the DIM (e.g.
Cannizzo 1993, and references therein), a thermal instability in the outer
disk creates an inward propagating heat front. This front transforms the disk
from a cold (quiescent) state to a hot state. Because the UV flux is mainly
emitted close to the white dwarf, one expects a delay in its rise, a delay
equal to the time it takes the front to travel from the outer disk to the
white dwarf. In the DIM, however, the calculated travel time of the front is
much shorter than the observed UV-delay time (Pringle, Verbunt \& Wade 1986;
Cannizzo \& Kenyon 1987). Thus, in its standard form the model fails to
explain the UV-delay.

Two solutions have been proposed in order to rescue the DIM; both of them
invoke a central ``hole'' in the accretion disk. At the edge of such a hole,
an inward moving heat front would have to stop. The hole would then fill up
on a viscous time scale, which is much longer than the heat front propagation
time. Livio \& Pringle (1992) suggested a mechanism for creating such a
hole: they argued that at quiescent mass accretion rates, the magnetic field
of a weakly-magnetized white dwarf can disrupt the inner accretion disk. They
showed that such a model can reproduce the UV-delay observed in dwarf nova
outbursts. This model cannot apply to systems in which the accreting object
is a black hole, since a black hole cannot support a permanent magnetic field.

Meyer \& Meyer-Hofmeister (1994) proposed a different scheme for quiescent
accretion onto a white dwarf that also results in a central hole. They
invoke inefficient cooling in the disk's upper layers, which leads to the
formation of a hot corona and ultimately to the evaporation of the inner
disk. As a result, the inner accretion flow consists of a pure coronal
plasma. A similar solution for quiescent SXTs has been independently
proposed by NMY. In both cases, the hot inner flow provides a natural
explanation for the hard X-ray emission observed in quiescent dwarf novae and
SXTs.

Whether a hole is created by magnetic fields or by evaporation, the effect on
the outburst of a dwarf nova is similar: when the heat front arrives at the
inner edge of the truncated disk it cannot propagate any further; the
(surface) density contrast slowly fills up the hole on a viscous time scale,
thereby producing the required delay of the UV outburst.

Below we show that the observations of the April 1996 outburst of \J16 imply
the presence of a two-component accretion flow in this system, and that the
parameters deduced from observations agree very well with a model comprised
of an outer cold disk and an inner hot ADAF, as proposed by NBM.

\subsection{The X-Ray Delay in the Outburst of GRO J1655--40}

We assume that the April 1996 outburst of \J16 was triggered by a dwarf-nova
type instability in a standard cold disk, which extends from the transition
radius $R_{\rm tr}$ to some outer radius $R_{\rm out}$. This disk is most
probably marginally stable with respect to this instability (e.g. Lasota et
al. 1996), or it may be globally unstable (see below). Once the instability
is triggered, the resulting heat front propagates with the speed $v_{\rm f}
\approx \alpha c_{\rm s}$ (Meyer 1984), where $c_{\rm s}$ is the
equatorial-plane sound speed in the hot phase. Thus, the time it takes for
the front to travel a distance $R_{\rm out}$ is
\begin{equation}
t_{\rm f} \approx {R_{\rm out} \over \alpha c_{\rm s}} \sim 2.8 \;
\alpha^{-1} R_{10} T_4^{-1/2} \rm \; hr,
\end{equation}
where $T_4 = (T/10^4 K)$ is the central disk temperature, and $R_{10}$ the
radius in units of $10^{10}$ cm. If the instability starts at a sufficiently
large radius, the front may take up to a day to reach the inner edge of the
disk; this has been proposed by ORBM as the origin of the observed delay
between the I, R, V and B light curves. Although very tempting, this
explanation suffers from the fact that for any reasonable values of the
radius and the primary mass, the effective temperature jumps to 10,000 --
11,000 K on a thermal time scale at the point where the instability sets in.
At these temperatures, B -- V $\sim$ 0, V -- I $\sim$ 0, and the B, V and I
magnitudes should increase simultaneously. Dilution of the disk light by the
ADAF and secondary light does not alter this conclusion. Thus the disk
instability model cannot easily account for a one day delay at optical/IR
wavelengths.

It is interesting to note that the observations of ORBM do not quite cover
the very initial rise of the outburst, since their first observation shows
that the R, V and I fluxes have already risen, whereas the B flux remains at
its quiescent value. This indeed implies the existence of a delay; however,
the value obtained by ORBM assumes a linear extrapolation, which may
overestimate the delay for two reasons. The initial rise could be very non
linear, with a sharp increase of the optical light from the disk, in which
case the delay could be only a few hours. Another possibility is that the
mass transfer rate from the secondary may have significantly increased since
the previous observation of the source in quiescence, one month prior to the
outburst. In fact, it is quite plausible that such an increase could have
triggered the thermal instability that caused the April 1996 outburst. In
any case, such an increase in the mass transfer rate would not show up in the
B-band because the quiescent temperature of the system does not exceed 6500 K.

When the heat front arrives at the transition radius $R_{\rm tr}$ where the
dense (cold) disk ends, it cannot propagate any further towards the black
hole; however, the resulting (surface) density contrast will propagate inward
due to viscosity. The speed at which the density ``front'' propagates is
\begin{equation}
v_{\rm visc} = {\nu \over w},
\end{equation}
where $\nu$ is the kinematical viscosity ($\nu=\alpha c_{\rm s} H$) and $w$
is the scale of the density gradient. The density contrast will travel a
distance $R_{\rm tr}$ in a time
\begin{equation}
t_{\rm vis} = {R_{\rm tr}\over v_{\rm visc}}.
\end{equation}
The width $w$ can be written as (see e.g. Cannizzo 1996)
\begin{equation}
 w = \beta (HR)^{1/2},
\end{equation}
where we expect $\beta$ to be $\lesssim 1$.

If we identify the observed X-ray delay with t$_{\rm vis}$ we can estimate
the transition radius as
\begin{equation}
r_{\rm tr} \approx 3.6 \times 10^4 \alpha^{4/3} t_{\rm X,d}^{4/3} m_1^{-4/3}
\beta^{-4/3} T_4,
\label{rtr}
\end{equation}
where $r_{\rm tr}= R_{\rm tr}/R_{\rm S}$, $R_{\rm S} = 2 G M/c^2$, $m_1 =
M/{\rm M_\odot}$ is the mass of the central black hole, and $t_{\rm X,d}$ is
the X-ray delay time in days. (Here and elsewhere we use the symbols $R$ and
$r$ to refer to the radius in physical units and Schwarzschild units,
respectively.) For $m_1=7$, $t_{\rm X,d}\approx 5$ and $\alpha=0.3$, one
obtains $r_{\rm tr} \approx 4.6 \times 10^3 \beta^{-4/3} T_4$ which shows
that the transition between the hot ADAF and the cold outer disk occurs at
$r_{\rm tr} \sim 10^4$. Remarkably, this same value of the transition radius
was deduced for models of two BHT by NBM in a completely independent way.
Furthermore, in the next section we show that $r_{\rm tr} \sim 10^4$ is close
to the radius at which the outer disk becomes unstable to the dwarf-nova
instability.

The outer disk radius in \J16 is
\begin{equation}
r_{\rm out} \approx 7.3\times 10^5 m_1^{-2/3} P_{60}^{2/3} \approx2\times10^5,
\end{equation}
where $P_{60}=P_{\rm orb}/60 {\rm h}$ is the orbital period.

\section{ADAF Plus Cold Disk Model for \J16 in Quiescence}

We have seen in the previous section that the X-ray delay in the outburst of
\J16 requires that the quiescent state of the system must consist of a
two-zone flow. The thin accretion disk can extend only down to a transition
radius $r_{tr}$ which has to be greater than a few thousand Schwarzschild
radii. Inside this radius, the flow must either be absent or must have a
much lower density than in the thin disk. This picture is very similar to
the two-zone model proposed by NMY and NBM for fitting the spectral data of
V404 Cyg and A0620--00 in quiescence; in that model, the flow inside the
transition radius consists of an extremely hot two-temperature ADAF. Here we
use the ADAF plus thin disk model to fit the spectral data on \J16 in
quiescence and thereby estimate some key parameters of the quiescent
accretion flow.

We first select a set of system parameters for \J16 that we use as inputs to
our models for the source. These include the black hole mass, the binary
inclination, the distance, and the velocity at the inner edge of the outer
thin accretion disk. Second, we summarize the multiwavelength data (X-ray,
optical and radio) that we use to constrain our models of the quiescent
state. Finally, we use the spectral data to constrain the remaining
parameters of the model.

The mass of the black hole in \J16 and the inclination of the system are very
well determined: we adopt M$_1 = 7 $~M$_{\odot}$ and $i=70^o$ (OB). In
\S2.2, we estimated the outer radius of the thin accretion disk to be $r_{\rm
out}\approx2\times10^5$; in the spectral models we take $\log(r_{\rm
out})=5.0$. This parameter does not need to be determined very accurately
since it has very little effect on the results. Based on studies of the
radio jets, we adopt a distance of D = 3.2 kpc (Hjellming \& Rupen 1995).

Dynamical and geometrical information about the thin accretion disk can be
obtained from studies of the broad, double-peaked Balmer lines (Smak 1981;
Horne \& Marsh 1986), and from this the transition radius $r_{\rm tr}$ can be
constrained. Of interest here is $v_{\rm in}$, the projected velocity at the
inner edge of the thin accretion disk. Estimates of this velocity have been
inferred for several SXTs from orbit-averaged profiles of the H$_{\alpha}$
emission line (see NMY, and references therein).

\J16 is a difficult case because of the strong H$_{\alpha}$ absorption line
present in the spectrum of the F subgiant secondary, and because of the
relative brightness of the secondary. In order to obtain a useful spectrum
at H$_{\alpha}$, we formed a sum of 73 spectra that had been collected over a
wide range of orbital phase in 1995 April-May when the system was near
quiescence (Bailyn et al. 1995), and then subtracted the spectrum of the
best-fitting F5IV star (Orosz 1996). In this way we measured the width of
the H$_{\alpha}$ emission line and estimated $v_{\rm in}$ $\geq$ 1045 km
s$^{-1}$. The inner edge of the thin disk, or equivalently the transition
radius $r_{\rm tr}$ between the thin disk and the ADAF, is then estimated to
be $r_{\rm tr}=(c\sin i/v_{\rm in})^2/2\leq3.6\times10^4$ (NBM). We present
below models corresponding to a range of values of $\log(r_{\rm tr})$
consistent with this constraint.

\begin{figure}
\plotone{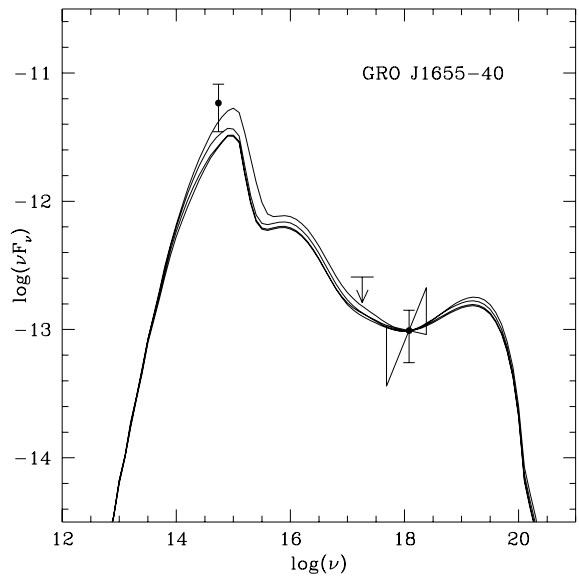}
\caption{Quiescent spectrum of the non-stellar component of \J16.
The dot and error bar on the left represent the estimated V band flux of the
quiescent accretion flow in \J16, the arrow shows the upper limit on the soft
X-ray flux as measured with ROSAT, and the dot on the right with error bar
and ``bow-tie'' indicates the ASCA constraint on the X-ray flux and spectral
index. The solid lines represent model spectra corresponding to an accretion
flow consisting of an inner ADAF plus an outer thin disk. In each case, the
mass accretion rate has been adjusted to fit the ASCA X-ray flux. From below
(in the optical band), the four models have transition radii, $\log(r_{\rm
tr})=4.5, ~4.0, ~3.5, ~3.0$, respectively.}
\end{figure}

\begin{deluxetable}{crrrlc}
\small
\tablewidth{0pt}
\tablecaption{Quiescent Spectrum of the Non-stellar Component of GRO J1655--40}
\tablehead
{
\colhead{Entry} & \colhead{Wavelength} & \colhead{log${\nu}$} & 
   \colhead{log($\nu$ F$_{\nu})^{a}$} & \colhead{Observatory} & 
   \colhead{Reference}\\
Number & \colhead{($\AA$)} & \colhead{(Hz)} & \colhead{(ergs cm$^{-2}$ s$^{-1}$
Hz$^{-1}$)} & \colhead{}&\colhead{}\\ 
}
\startdata
1 & 0.05 & 19.818$^{b}$ & $<$-9.49 & CGRO/OSSE & 1\\
2 & 0.17 & 19.258$^{b}$ & $<$-9.44 & CGRO/BATSE & 1\\
3 & 2.3~ & 18.123$^{c}$ & -12.96 & ASCA & 1\\
M
4 & 16.6~ & 17.258$^{b}$ & $<$-12.59 & ROSAT/HRI & 1\\
5 & 5500~~~ & 14.736~ & -11.24 & CTIO & 2 \\
6 & 3.6x10$^{8}$ & 9.924~ & $<$-16.38 & VLA & 1\\
7 & 6.1x10$^{8}$ & 9.690~ & $<$-16.61 & VLA & 1\\ \hline
\enddata
\tablenotetext{a}{Flux limits are at the 3$\sigma$ level of confidence.}
\tablenotetext{b}{Central frequency computed assuming a Crab-like
spectrum with energy index $\alpha_{E} = 1.1$ (see NBM).}
\tablenotetext{c}{Central frequency computed assuming $\alpha_{E}
= 0.5$ (see Sect.~3).}
\tablerefs{(1) Robinson et al. (1997); (2) Orosz \& Bailyn 1997}
\end{deluxetable}

The multi-wavelength data, $\nu F_\nu$ vs. $\nu$, are summarized in Table 1.
To derive the optical flux (entry 5), it was necessary to subtract a large
stellar component. We assumed that the residual (non-stellar) component
contributed 5 $\pm$ 2\% of the total light (OB); for the total optical flux
from the system we used the apparent magnitude and reddening given in OB.
The optical flux in the V band and its error bar are shown in Fig. 2.

Apart from the optical and BATSE data (entry 2), all the remaining data in
Table 1 were obtained during an intensive campaign of observations of \J16 in
quiescence conducted in March 1996 by Craig Robinson and his collaborators
(Robinson et al. 1997). All of the upper limits in Table 1 and below are
quoted at the 3 sigma level of confidence. The OSSE flux limit (entry 1)
corresponds to an intensity upper limit of 40 mCrab (100--600 keV). This is
off-scale in Fig. 2 and is not plotted. \J16 was detected by ASCA (entry 3)
at an (unabsorbed) flux level of F$_{\rm x}$ (2--10 keV) = (1.6 $\pm$ 0.7)
$\times 10^{-13}$ erg cm$^{-2}$ s$^{-1}$; the photon power-law index
($\alpha$) is estimated to be 1.5 $\pm$ 0.6 (Robinson et al. 1997). Both the
flux (with its error bar) as well as the allowed range of spectral slope are
indicated in Fig. 2. A stringent flux limit was obtained using the ROSAT HRI
detector (entry 4): F$_{\rm x}$~(0.2--2 keV)~$<$~6.1 $\times 10^{-13}$ erg
cm$^{-2}$ s$^{-1}$ ($3\sigma$), assuming $\alpha$ = 2.1 and N$_{\rm H}$=$5
\times 10^{21}$ cm$^{-2}$ (Robinson et al. 1997). This upper limit is
indicated by the arrow in Fig. 2. Finally, the VLA limits on \J16 (entries 6
\& 7) correspond to a flux limit of 0.5 mJy at both 4.9 GHz and 8.4 GHz (not
shown on Fig. 2).

The BATSE limit (entry 2) is an average over the period 19--30 April 1996 and
corresponds to a photon flux (20--200 keV) of $-0.0001\pm 0.0024$ ph cm$^{-2}$
s$^{-1}$ (assuming $\alpha$ = 2.8), or an intensity limit of 26 mCrab. This
is consistent with the OSSE limit. Recall, however, that the inferred start
time of the X-ray rise at 2--12 keV is 25.4 $\pm$ 0.8 April 1996 (Orosz et al.
1997). So the BATSE limit includes a $\sim$ 5--day period when the 2--12 keV
X-ray outburst was underway.

We have attempted to fit the quiescent spectral data on \J16 using an ADAF
plus thin disk model, analogous to the models described in NBM. The solid
lines in Fig. 2 represent four models, where each model consists of a pure
ADAF for radii $\log(r)<\log(r_{\rm tr})$, and a thin accretion disk in the
radius range $\log(r_{\rm tr})\leq\log(r)\leq\log(r_{\rm out})$. The models
correspond to $\log(r_{\rm tr})=4.5, ~4.0, ~3.5$ and 3.0, respectively. The
transition from the thin disk to the ADAF in these models occurs via
evaporation into a corona, as described in NBM. The models assume
equipartition between gas and magnetic pressure ($\beta=0.5$ in the notation
of NMY and NBM) and the viscosity parameter is taken to be $\alpha=0.3$ in
the ADAF region. It is assumed that a fraction 0.001 of the viscous energy
directly heats the electrons in the ADAF (and the corona) while the remaining
0.999 of the energy goes initially into the ions (i.e. $\delta=0.001$, see
NBM for details). The shape and normalization of the computed spectra are
quite insensitive to the values chosen for $r_{\rm tr}$, $\beta$, $\alpha$
and $\delta$ (see Figures 3--5 in NBM).

In each model, only one parameter has been adjusted, namely the mass
accretion rate. This has been optimized such that the model flux in the ASCA
band agrees with the observed flux. Despite the large range of $r_{\rm tr}$
covered by the four models, the mass accretion rates vary very little from
one model to another; in Eddington units, the accretion rates range from
0.0034 to 0.0037, which correspond to physical accretion rates of $\dot
M=(3.4--3.7)\times10^{16} ~{\rm g\,s^{-1}}$. Thus, the spectral models
constrain the $\dot M$ of \J16 in quiescence quite well. Technically, the
models determine only the parameter combination $\dot M/\alpha$ and so $\dot
M$ depends on a knowledge of $\alpha$. However, the value of $\alpha$ in
ADAFs is fairly well-constrained by the various studies done to date (Narayan
1997), and is unlikely to vary by more than a factor $\sim 3$ either way from
the value we have assumed, $\alpha=0.3$. This suggests that the above
estimate of $\dot M$ in the ADAF is good to about a factor $\sim3$.

The models shown in Fig. 2 are consistent with all the measurements available
at this time, including the OSSE and VLA limits (which are not shown in Fig.
2). Note, in particular, that the models fit the observed optical flux,
predict the correct slope in the ASCA band, and lie below the ROSAT flux
limit.

A rather obvious point is that the quiescent data are incompatible with any
model which is based only on a thin accretion disk. The ROSAT and ASCA data
clearly indicate that (1) the X-ray flux of \J16 in quiescence lies below the
optical flux, and (2) the X-ray spectrum is quite hard, with a photon index
$<2.7$ (2 $\sigma$). A thin accretion disk model, with either a constant or
variable $\dot M$ as a function of radius, cannot possibly reproduce such a
spectrum. Thus, \J16 is similar to A0620--00 and V404 Cyg (see NMY and NBM),
where again the quiescent spectra are found to be inconsistent with a pure
thin disk model but are fitted well with an ADAF plus thin disk model.

Although the spectral fit does not help us to determine $r_{\rm tr}$, it is
possible to obtain a fairly strong constraint on $r_{\rm tr}$ by considering
the stability of the outer thin disk (Lasota et al. 1996, NBM).
Specifically, the outer disk will be unstable to the dwarf nova instability
if it has an effective temperature greater than about 5000 K, and therefore
the quiescent disk cannot exceed this temperature at any radius. The four
models presented in Fig. 2, with $\log(r_{\rm tr})=4.5, ~4.0, ~3.5, ~3.0$,
have maximum effective temperatures in their disks of $T_{\rm max}=1700,
~3700,$ $~8400,~20000$ Kelvin respectively. The requirement $T_{\rm
max}<5000$ K thus provides the constraint $\log(r_{\rm tr})>3.7$, or $r_{\rm
tr}>5000$. Just prior to outburst, we expect the thin disk to be very close
to the limiting value of $T_{\rm max}$. We therefore estimate that \J16 had
its transition radius at $r_{\rm tr}\sim5000$, or $R_{\rm tr}\sim10^{10}$ cm,
at the time of the April 1996 outburst.

\section{Instability of the Outer Disk}

In this section, we present numerical simulations of the dwarf nova
instability in the outer disk of \J16. We compare the results with
observations of the early stages of the outburst, paying particular attention
to the 6--day delay between the optical and X-ray outburst. We also compare
the mass accretion rate implied by the outburst calculations with the
quiescent $\dot M$ determined independently in the previous section. The
calculations presented here have been done with the code described in Hameury
et al. (1997).

In the following we assume that the mass transfer rate from the companion
star, i.e. the accretion rate at the outer rim of the accretion disk, has the
value given by OB, viz. $\dot M_{\rm transfer}=2\times 10^{17} ~{\rm
g\,s^{-1}}$. For this value of $\dot M_{\rm transfer}$, the outer cold disk
is unstable to the dwarf nova instability (see e.g. Ludwig et al. 1994), and
so we are guaranteed that the code will produce an outburst.

We assume that the transition between the outer thin accretion disk and the
ADAF is due to evaporation into a corona which gradually erodes matter in the
disk as the cold inflowing material approaches the transition radius; such a
model has been proposed for dwarf novae by Meyer \& Meyer-Hofmeister (1994;
see also NMY \& NBM). We have estimated the evaporation rate by the
following approximate method. Narayan \& Yi (1995) have derived that the
maximum allowable mass transfer rate in the ADAF at small radii is $\dot
m_{\rm ADAF,max} = 0.3 \alpha^2$ (in Eddington units) and that $\dot m_{\rm
ADAF,max}$ decreases at large radii ($r>10^3$) (see also Abramowicz et al.
1995). Assuming that the mass transfer rate within the inner ADAF is equal
to the maximum, and using M$_{1} = 7 \Msun$, we adopt the the following
approximate prescription for evaporation:
\begin{equation}
\dot{M}_{\rm ev} = {2.8 \times 10^{17} \over (1 + K R_{\rm
tr,10}^2)} \;\; \rm g\; s^{-1} ,\qquad R_{10}\geq R_{\rm tr,10},
\end{equation}
where $K$ is a constant which is adjusted so as to give the required value of
the transition radius, and $R_{\rm tr,10}$ is the transition radius in units
of 10$^{10}$ cm. The local surface density evaporation rate is then related
simply to the derivative of $\dot{M}_{\rm ev}$ with respect to $R$, i.e.
\begin{equation}
\dot{\Sigma}_{\rm ev} = {1\over2\pi R}{d\dot M_{\rm ev}\over dR}=
{9 \times 10^{-4} K \over (1 + K R_{10}^2)^2} \;\; \rm
g\;s^{-1}cm^{-2} .
\end{equation}
This prescription for the evaporation is numerically close to the formula
given by Meyer \& Meyer-Hofmeister (1994).

According to the dwarf nova DIM, the accretion rate in a quiescent disk must
satisfy
\begin{equation} \dot M(r) <
\dot M_{\rm crit}(r) = 9.6 \times 10^3 m_1^{1.73} r^{2.6} \ {\rm
g\ s^{-1}},
\end{equation}
where we have taken $\dot M_{\rm crit}$ from Ludwig et al. (1994). The disk
first becomes unstable at its inner edge when $\dot M$ in the disk reaches
the critical value near the transition radius. This triggers an inside-out
outburst. Since most of the mass evaporation occurs close to the transition
radius, the condition for the outburst is equivalent to the requirement
$\dot{M}_{\rm ev}=\dot{M}_{\rm crit}$. For $K R_{\rm tr,10}^2 > 1$, we then
find
\begin{equation}
R_{\rm tr} = 3.3 \times 10^{10} K^{-0.43} m_1^{0.19} \; \rm cm.
\end{equation}
As we showed in the previous section, \J16 in quiescence requires a
transition radius $\sim10^{10}$ cm, which means that we require a value of
$K$ of order a few. In the detailed calculations presented below we have
selected $K=5$, which gives $R_{tr}=10^{10}$ cm for the quiescent model just
before outburst. In this model, the mass transfer rate feeding the ADAF
prior to the onset of the instability is found to be $4.6\times10^{16} ~{\rm
g\,s^{-1}}$, which is in excellent agreement with the $\dot M$ in the ADAF
estimated in Sect. 3 ($\dot M\sim3.5\times10^{16}~{\rm g\,s^{-1}}$ for
$\alpha_{\rm ADAF}=0.3$) by fitting the spectrum of \J16 in quiescence.

In the presence of evaporation the usual disk equation for mass conservation
has to be modified as follows:
\begin{equation}
{\partial \Sigma \over \partial t} + \dot{\Sigma}_{\rm ev} = -{1 \over
R}{\partial \over \partial R} (R \Sigma v_{\rm R}),
\end{equation}
where $\Sigma$ is the surface column density in the disk, and $v_{\rm R}$ is
the radial velocity. Because the evaporation law is independent of $\Sigma$,
evaporation results in a disk which is sharply cut off at the transition
radius $R_{\rm tr}$. We thus use as an inner boundary condition the relation
\begin{equation}
{\dot M_{\rm disk}}(R_{\rm tr}) = 0.
\end{equation}
Once the outburst begins, the transition radius moves in and reaches $R_{\rm
in}$, the inner edge of the grid. When this happens, in order to avoid
numerical complications, we switch to the standard inner boundary condition,
$R = R_{\rm in}$.

In the calculations presented here, the disk inclination was taken to be
$70^o$, the outer radius of the disk was taken to be $4\times10^{11}$ cm, and
the inner boundary of the grid was set at $R_{\rm in}=4\times10^8$ cm. Our
choice of the inner boundary corresponds to $R_{\rm in}=194R_{\rm S}$ rather
than $3R_{\rm S}$, but this is merely for numerical convenience and does not
affect any of the results presented here. Once the outburst gets underway
and the thin disk extends inward to $R_{\rm in}$, the time it needs to travel
the additional distance to the black hole is quite short compared to the time
it took to move from $R=R_{\rm tr}=10^{10}$ cm down to $R=R_{\rm
in}=4\times10^8$ cm. Therefore, very little error is made by truncating the
numerical simulation at $R_{\rm in}$. The disk viscosity parameter $\alpha$
in the simulations varies between 0.035 on the cool branch and 0.15 on the
hot branch.

\begin{figure}
\epsscale{0.90}
\plotone{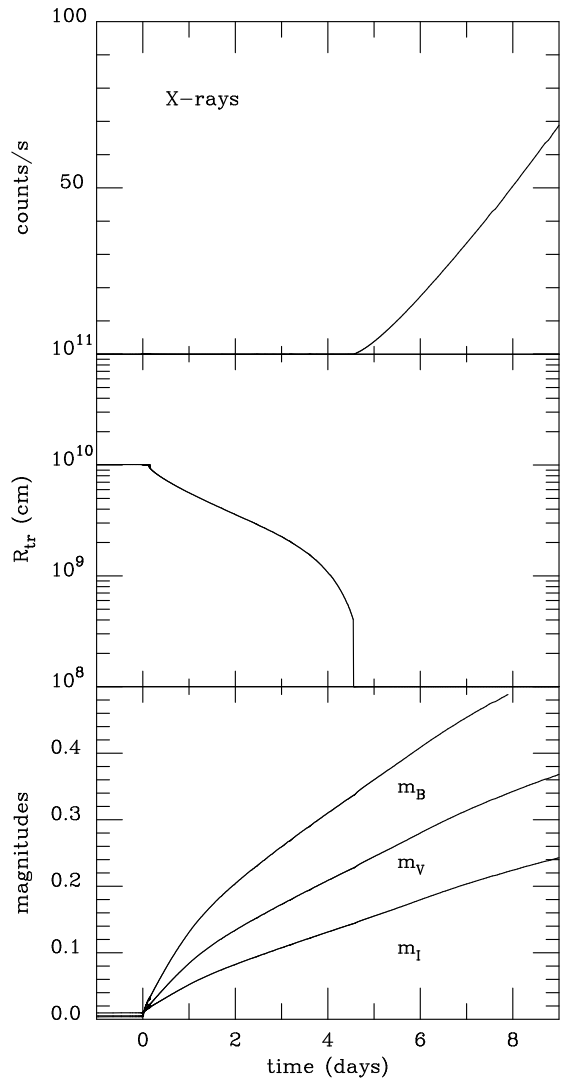}
\caption{Results obtained by modeling the outburst in \J16 by
means of a dwarf-nova like instability in the outer thin disk. The upper
panel shows the predicted X-ray flux expected to be observed by the RXTE ASM,
the second panel shows the variation of the transition radius $r_{\rm tr}$
with time and the lower panel shows the variations of the optical B, V, and I
magnitudes. In order to avoid numerical difficulties, the inner edge of the
numerical grid was fixed at $4 \times 10^8$ cm; however, when $r_{\rm tr}$
reaches this value, it is set to $3 R_{\rm S}$ in the plot, which is more
realistic.}
\end{figure}

Figure 3 displays the initial phases of an outburst seen in the numerical
calculations. In this run, several outbursts have already occurred, so that
the initial assumed density profile in the disc has been relaxed. The bottom
panel shows the magnitude variations $-\Delta m = -m + m_{0}$ in the B, V and
I bands, where $m_{0}$ accounts for the presence of diluting light
originating essentially from the secondary (which dominates over the ADAF).
For simplicity, we have assumed that $m_0$ is constant, and corresponds to a
48 L$_\odot$ secondary with an effective temperature of 6500 K (OB). During
quiescence, the disk is extremely faint---fainter than both the secondary and
the ADAF---and contributes less than 1 \% of the total light; however, in
outburst, its optical luminosity, although still smaller than that of the
secondary, dominates the ADAF, justifying our assumption that $m_0$ is
constant.

The B, V, and I magnitudes decrease simultaneously in the calculations; this
is independent of the magnitude of the diluting light $m_0$. The slopes
however are directly related to $m_0$: for large diluting fluxes, the
logarithm appearing in the definition of the magnitude can be linearized, and
one has ($L(t) \ll L_0$)
\begin{equation}
-\Delta m = 1.09 {L(t) \over L_0},
\end{equation}
where $L(t)$ is the disk luminosity and $L_0$ the luminosity of the companion
plus the ADAF. The faster rise in the B band is thus simply due to the fact
that B--V = 0 for a disk in the hot state, whereas most of the diluting light
comes from the secondary with B--V $\sim$ 0.5. This gives a factor $\sim$
1.5 between the slopes of the B and V magnitudes, as observed.

The middle panel in Fig. 3 shows the variation of the transition radius with
time. Once the outburst begins, the mass accretion rate increases
significantly and the evaporation is unable to keep up. The transition
radius therefore decreases, slowly at first and then more rapidly as the
characteristic viscous time decreases with decreasing radius. About 5 days
after the start of the outburst, $R_{\rm tr}$ moves down to $R_{\rm in}$, the
inner boundary of the numerical grid. At later times, we assume that
whatever accreted matter reaches $R_{\rm in}$ will continue down to the black
hole in the form of a thin disk, and we calculate the X-ray luminosity
accordingly.

The top panel in Fig. 3 shows the calculated X-ray light curve. This has
been computed assuming that the emission from the ADAF has an efficiency of
0.1\%, while the matter flowing through $R_{\rm in}$ in the thin disk has a
standard efficiency of 10\%. We assumed a conversion factor of
$5.3\times10^{35}$ ergs/count to relate ASM counts/s (2--12 kev) to X-ray
luminosity (ergs/s). This simulation shows that it takes about 5 days for
the transition radius to move from its initial value of 10$^{10}$ cm to
values small enough that X-rays can be emitted, in excellent agreement with
the optical to X-ray time delay observed in the April 1996 outburst of \J16
(ORBM). This explanation of the observed delay is the most outstanding
success of the present calculations.

The simulations predict that the various optical bands should go into
outburst simultaneously, whereas in the observations the outburst occurred
first in the I band, followed by the other bands in the order R, V, and B,
spread over a time range of about a day. The model does not reproduce the
delays in the optical bands, but, as explained in Sect. 2.2, this delay could
result from an earlier increase in the mass transfer rate that occurred less
than a month before the outburst, and could well have caused it. It is also
worth noting that our optical light curves deviate significantly from
linearity during the first two days, whereas the subsequent evolution shows
an almost linear variation of the optical magnitudes. This suggests that it
may be somewhat difficult observationally to identify the relative time of
outburst in various optical bands.

Both the quiescent mass transfer rate into the ADAF and the inner disk radius
are solely determined by the evaporation law. On the other hand, the X-ray
delay corresponds to the time it takes to rebuild a standard inner disk, and
is thus proportional to the mass of the disk, and inversely proportional to
the mass transfer rate at the transition radius. This mass transfer rate
depends essentially on $\alpha_{\rm h}/\alpha_{\rm c}$, the ratio of the
Shakura-Sunyaev viscosity parameter in the hot and cold states of the disk.
Therefore, an increase in $K$ results in a smaller transition radius and a
smaller ADAF luminosity; the corresponding shortening of the X-ray delay can
in turn be compensated for by taking a smaller $\alpha_{\rm h}$, i.e. by
increasing the viscous time.

The calculations described so far show that a two-component accretion flow
model consisting of an outer thin disk and an inner ADAF explains most of the
key observations of \J16. It explains the quiescent spectrum of the source
as well as the characteristics of the outburst, notably the X-ray delay. We
argued in \S3 that the quiescent spectrum of \J16 cannot be explained by a
pure thin disk. We now show that a pure thin disk model has difficulty in
reproducing the outburst observations. We consider two models for which the
thin disk extends down to $R_{\rm in}=4\times10^8$ cm.

\begin{figure}
\plotone{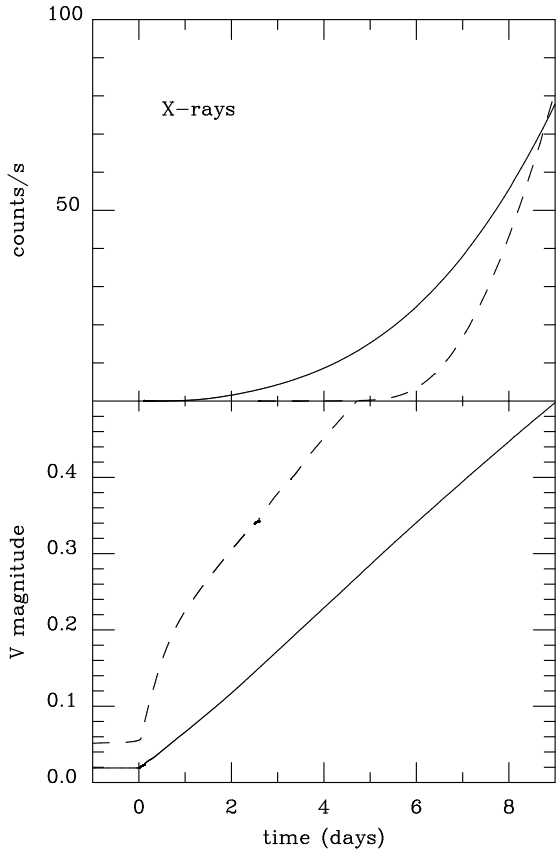}
\caption{Same as Fig. 3 but for outbursts in a thin cold disk
extending down to $4 \times 10^8$ cm. The solid curve corresponds to an
inside-out outburst; the dashed curves to an outside-in situation. In the
latter case, matter was artificially added at $8 \times 10^{10}$ cm to
trigger the outburst.}
\end{figure}

The first model describes an inside-out outburst (solid lines in Fig. 4).
The surface density in the outer parts of the accretion disk was adjusted
such as to reproduce the correct slope of the X-ray and optical light curves,
the viscosity being as in the previous model. For this reason, the
agreement between the predicted and observed light curves is very good,
better in fact that in the case of Fig. 3 for which such a fitting procedure
was not performed, since we considered a relaxed case (i.e. after several
outbursts) in order to minimize the number of free parameters. The X-ray
intensity, which increases simultaneously with the optical flux, is initially
quite faint. Thus, the X-ray delay depends on the sensitivity of the X-ray
detector; for example, the ASM would be unable to detect the X-rays during
the first 1--2 days, and therefore the model predicts an X-ray delay of this
order. The delay could be increased further by decreasing the viscosity
parameter, which would slow the rise of the X-ray intensity. However, this
would cause the rise in the optical flux to be unacceptably slow.

The second model (dashed lines in Fig. 4) describes an outside-in outburst.
In order to obtain the longest possible characteristic times, and therefore
the most optimistic scenario for the pure disk model, we started the outburst
as far out in the accretion disk as possible. In long period systems like
\J16, the outer disk is always cold and stable; therefore to trigger the
outburst we added some matter at $R= 8 \times 10^{10}$ cm. The outburst then
began at $R= 7 \times 10^{10}$ cm. The viscosity was chosen such as to
reproduce the observed X-ray delay ($\alpha_h = 0.10$): it takes 2.6 days for
the heat front to reach the inner edge of the disk, and an additional 3.1
days for the mass accretion rate to reach 10$^{16}$ g s$^{-1}$, the level at
which the X-ray flux becomes detectable. Thus, this model reproduces the
observed X-ray delay. However, the optical light curves are not in agreement
with observations since most of the disk reaches a hot state before the heat
front reaches the inner edge of the disk. Consequently, the optical flux
increases too rapidly, on a thermal time scale, and then more slowly, on a
viscous time scale. The observed optical light curve does not exhibit such a
prominent two-phase behaviour, nor does it show such a rapid optical
increase, features which are the signatures of an outside-in outburst.

It therefore appears difficult to reconcile the observations with a pure disk
model. This might not be an insuperable difficulty, but it would most
probably require making ad-hoc assumptions about the density profile in the
disk and about the viscosity. Even if a candidate model could be contrived,
one would also require an explanation for the quiescent X-ray flux that does
not invoke accretion, since the whole disk must be in the low state and
therefore the mass transfer rate has to be less than $5 \times 10^6$ g
s$^{-1}$ at the last stable orbit. Such a small rate of mass transfer
corresponds to a luminosity of less than $5 \times 10^{26}$ erg s$^{-1}$,
almost six orders of magnitudes below the observed quiescent luminosity.
Together, the difficulty of building a viable model plus the near
impossibility of explaining the X-ray flux in quiescence rule strongly
against the pure disk model.

\section{Conclusions}

We have shown that both the X-ray spectrum observed in quiescence and the
6--day delay between the optical rise and the X-ray outburst in \J16 imply
that the accretion disk does not extend in its ``standard'' form all the way
down to the last stable orbit; instead the inner part of the accretion flow
is a hot ADAF. It is this ADAF region that is responsible for the X-ray
emission detected by ASCA in quiescence. The outer parts of the disk, located
at distances larger than about 10$^{10}$ cm in quiescence, are cold and
subject to the same thermal and viscous instability as in dwarf novae. After
the instability has been triggered, a heat front propagates inward. When
this heat front reaches the transition radius, the divide between the thick
disk and the ADAF region, it cannot propagate in further until it rebuilds
the inner part of the disk on a viscous time scale (about a week). Because
the efficiency of energy release in the ADAF region is very low, the X-ray
outburst starts only when the dense parts of the disk can penetrate far
enough in, close to the marginally stable orbit, to allow an efficient
transformation of gravitational energy into radiation.

We have also shown that both in quiescence and during the initial outburst
the observed properties of \J16 are not consistent with a pure locally cooled
thin accretion disk without an ADAF component. If one wishes to invoke such
a model, one has to (1) accept that the quiescent X-ray emission is not
linked to accretion, and (2) tune the viscosity and the initial surface
density of the disk with some care.

\bigskip
\bigskip

We are grateful to C. Robinson for making the results of his multiwavelength
campaign of observations of GRO J1655--40 available to us prior to
publication, and to J. Orosz for several fruitful discussions on his optical
studies of GRO J1655--40 and for providing us the required data to construct
Fig. 1. Partial support for J.E.M. was provided by the Smithsonian
Institution Scholarly Studies Program. R.N. was supported in part by NASA
grant NAG 5--2837.

\clearpage

\end{document}